\documentclass{article}  
\usepackage{breckenr2005}
\usepackage{graphicx}
\frompage{000} \topage{000}                                              
\def\rightmark{Puzzling Scaling in HBT}
\def\leftmark{M.A. Lisa (STAR Collaboration)}

\newcommand {\Alm} {{\mbox{$A_{\lambda,\mu}$}}}
\newcommand {\AlmQ} {{\mbox{$A_{\lambda,\mu}(Q)$}}}

\title{AA versus pp (\& dA):
A Puzzling Scaling in HBT@RHIC}

\authors{
{Z.~Chajecki$^a$, T.D.~Gutierrez$^b$,
M.A.~Lisa$^a,*$,
and M.~L\'{o}pez-Noriega$^a$, for the STAR Collaboration
}\\[2.812mm]
{\normalsize
\hspace*{-8pt}$^a$ Ohio State University, Columbus, OH 43210, USA\\[0.2ex] 
\hspace*{-8pt}$b$ Lawrence Berkeley National Lab, Berkeley, CA, USA\\[0.2ex] 
}}
 
\abstract{A nontrivial space-time structure of the hot system created at RHIC is
{\it the} defining aspect of the physics of relativistic heavy ion collisions.
Femtoscopy through pion intensity interferometry provides direct access to the
dynamic geometric substructure of the freeze-out stage, and appears to confirm
source evolution via collective flow.  Since flow is usually considered to be
a {\it bulk} phenomenon, it is surpring to find a simple scaling in preliminary
HBT radii from p+p, d+Au and Au+Au collisions.
Investigating the light system data in detail, we discuss 
a new way to visualize the ``fine structure'' of 3D 
correlation functions and the potential importance of long range correlations.}

 
\begin{document}
 
\setcounter{page}{1}
\maketitle

\section{Introduction}\label{sec:introduction}
\vspace*{-2.0mm}

A central goal in the study of relativistic heavy ion collisions is to probe an
equilibrated {\it bulk} system of partonic matter.  Copious results obtained at
the Relativistic Heavy Ion Collider (RHIC) indicate the early ideas of the creation
of a gas of quasi-free quarks and gluons (the Quark Gluon Plasma or QGP) need to be
replaced a non-hadronic ``liquid''~\cite{ShuryakPerfect} of strongly interacting color matter
(the sQGP~\cite{GyulassySQGP}), in which the degrees of freedom are composite
quasi-particles~\cite{ShuryakPerfect}.  Understanding the nature of this matter at
the most extreme conditions is of fundamental interest.

The strongly interacting nature of the sQGP is indicated by very strong collective flow~\cite{flow}
and energy loss of hard-scattered partons; these are indisputably the most trumpeted observations at
RHIC.  However, these exciting probes deal exclusively with momentum space, and thus fail to probe a crucial
part of the story.  

As the apparatus, terminology, and techniques of modern heavy ion physics 
resemble more and more those of particle physics, it is worthwhile to
remember that the distinguishing feature of our field is a focus on
non-trivial space-time structure.  The defining aspect of the new state
of matter hoped for at RHIC is a change in the {\it space-time} length
scales characterizing the system and color correlations.  Further, this
new phase of matter can only be generated in collisions of ``large'' systems.
Geometric features of the nuclear overlap in the initial state
(especially at finite impact paramter) drive the further evolution of
the system.  In the intermediate state (during or after thermalization),
collective bulk behaviour develops, generating non-trivial correlations
in momentum and coordinate space.  Further, in the final state, anomolously
large spatio-temporal scales may signal the the phase change from the
color-deconfined to confined (hadronic) state.  Clearly, space-time
aspects of the system evolution are of fundamental importance to understanding
the system at RHIC.  In such a context, disappointing indeed are recent
tendencies~\cite{GyulassySQGP} to dismiss measurements (annoying in that they disagree with
calculations) with strong connection to the space-time structure of heavy ion collisions.

Two-particle correlations at low relative momentum (for historical reasons
often called ``HBT correlations'' when performed
with identical mesons) represent the most direct method to extract space-time scales from
a femtoscopic
dynamic 
source.
Pion HBT has been used extensively to extract the size~\cite{RHICHBT,STARHBTPRC}
and shape~\cite{STARasHBT} of the freeze-out configuration at RHIC.  In the
next Section, we very briefly review the status of our understanding HBT in heavy ion collisions.
This motivates the present study which compares small collision systems (p+p, d+Au, and peripheral
Au+Au) to the largest ones (central Au+Au).  Preliminary results suggest a surprising scaling
of length scales between these systems.  In an effort to understand the signal for the small
systems, we introduce a new representation of the correlation function which points to the
importance of long-range and not yet understood correlations.

\section{Comforting femtoscopic systematics}
\vspace*{-2.0mm}

Heavy ion physics observables are always determined by an interplay between various underlying processes.
Thus, extracting meaningful information from any observable (e.g. flow, spectra)
requires systematic study of the observable over a large range of conditions in which the relative
contribution between the competing physics changes.  The history of heavy ion physics is littered
with apparent understandings of underlying physics, based on agreement of theory with data under
very limited conditions.  But when the data is studied systematically (e.g. by varying collision
energy), the theory is seen to fail.  One hopes that we are not being similarly misled at RHIC, where
several successful theories are considered valid only at these energies, and only for relatively central collisions,
so that failure to reproduce data under different conditions is not considered a problem.

Perhaps even more than other observables, femtoscopic measurements need to be systematically
studied, since questions of ``puzzles'' in the space-time sector at RHIC have been raised.
Here, we touch on some of the main femtoscopic systematics at that AGS, SPS, and RHIC.  A
full discussion may be found in~\cite{ReviewPaper}.

\begin{figure}[t!]
 \begin{minipage}[t]{0.4\textwidth}
 {\centerline{\includegraphics[width=0.7\textwidth]{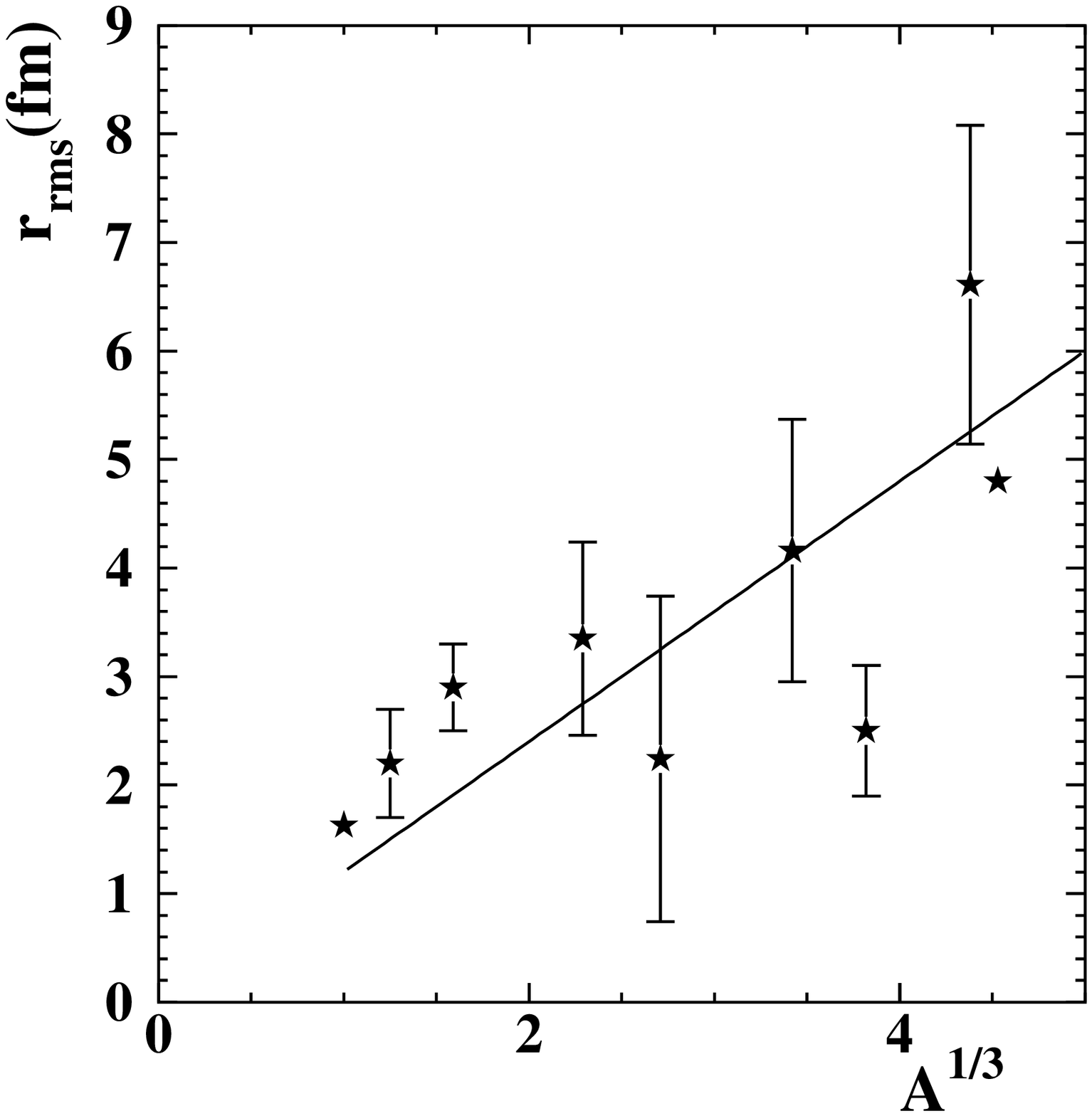}}}
     \caption{An indication of the state of femtoscopy 15 years ago.  Data from~\cite{Chacon} compiled in~\cite{Alexander}.
 }
     \label{fig:Status15YearsAgo}
 \end{minipage}
 \hspace{\fill}
 \begin{minipage}[t]{0.6\textwidth}
 {\centerline{\includegraphics[width=0.7\textwidth]{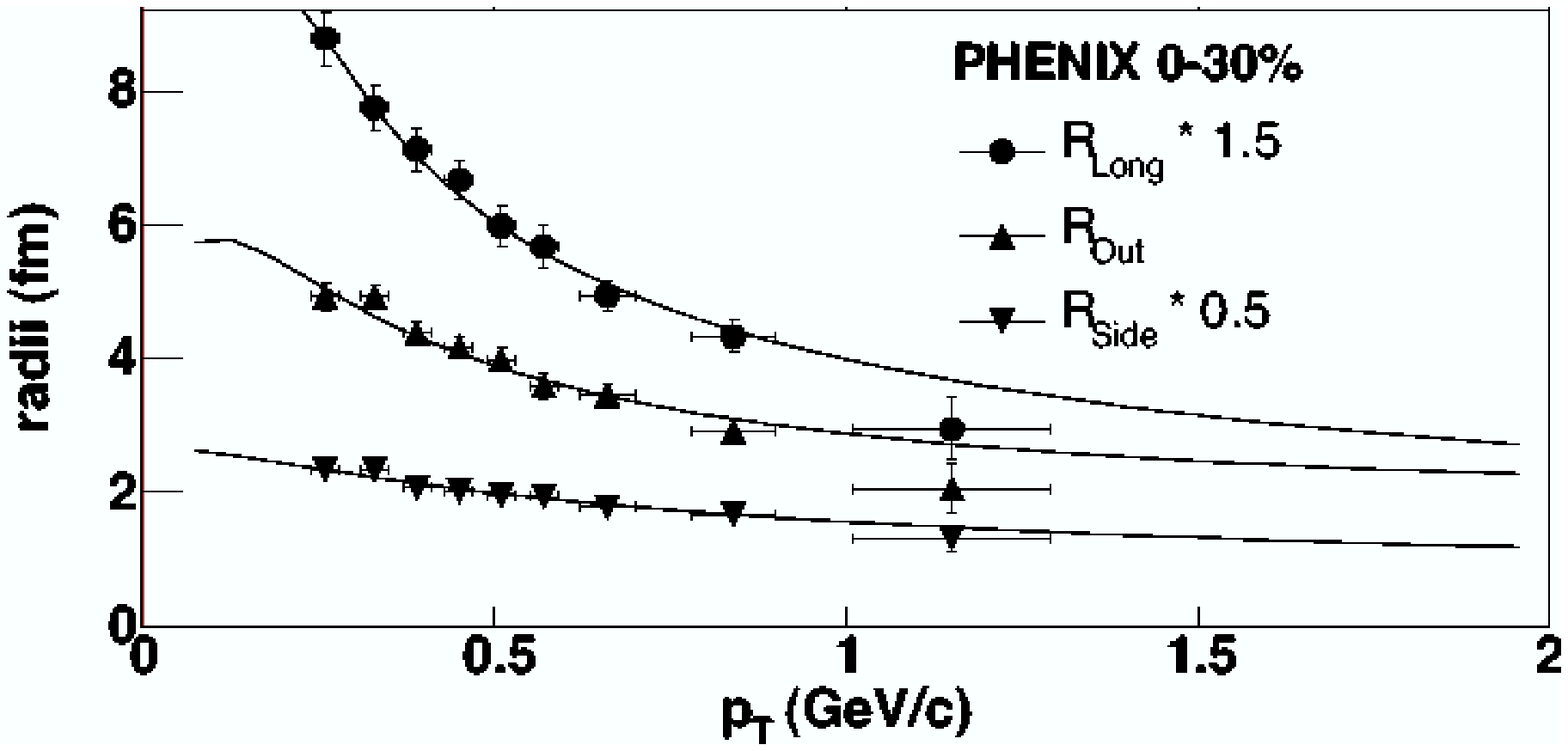}}}
     \caption{Pion HBT radii from central Au+Au collisions at RHIC, fitted~\cite{FabriceQM} with the Blast-wave model~\cite{RLBW}
 using flow and temperature parameters from momentum spectra.
 }
     \label{fig:RLBWHBT}
 \end{minipage}
 \end{figure}

To confirm the most basic premise of interferometry, it is important and comforting to note that
transverse (to the beam) length scales increase monotonically with the size of the colliding system.
Fifteen years ago, this was almost {\it all} that was known about HBT radii~\cite{Chacon}, and
even that with very large error bars.  Figure~\ref{fig:Status15YearsAgo} represents the status of
systematic femtoscopy in heavy ion physics prior to initial runs at AGS and SPS.
In the past decade-and-a-half, data and techniques have evolved tremendously, generating
an equally tremendous range of systematic femtoscopic studies.


The extracted transverse {\it shape} of the system at freeze-out also makes intuitive sense.  At AGS
energies, for which elliptic flow (preferred expansion in the reaction plane) is weakest, the freezeout
shape approximates that of the overlap region between the target and projectile.  At RHIC, strong elliptic
flow leads to a considerably rounder shape.

Kinematic systematics encode more detailed and more important
dynamically-generated geometric substructure.  Rapidity dependent femtoscopy
supports a freeze-out configuration which is approximately boost-invariant at midrapidity, underscoring
the role of longitudinal flow.  More interesting, however, are transverse collective flow components,
since they must be generated in the evolution of the hot system and provide some information on
thermalization.  The strong decrease of transverse HBT radii (from pions and other correlated particles)
with $m_T$ is believed to result directly from space-momentum correlations induced by flow.

\begin{figure}[t!]
 \begin{minipage}[t]{0.5\textwidth}
 {\centerline{\includegraphics[width=0.7\textwidth]{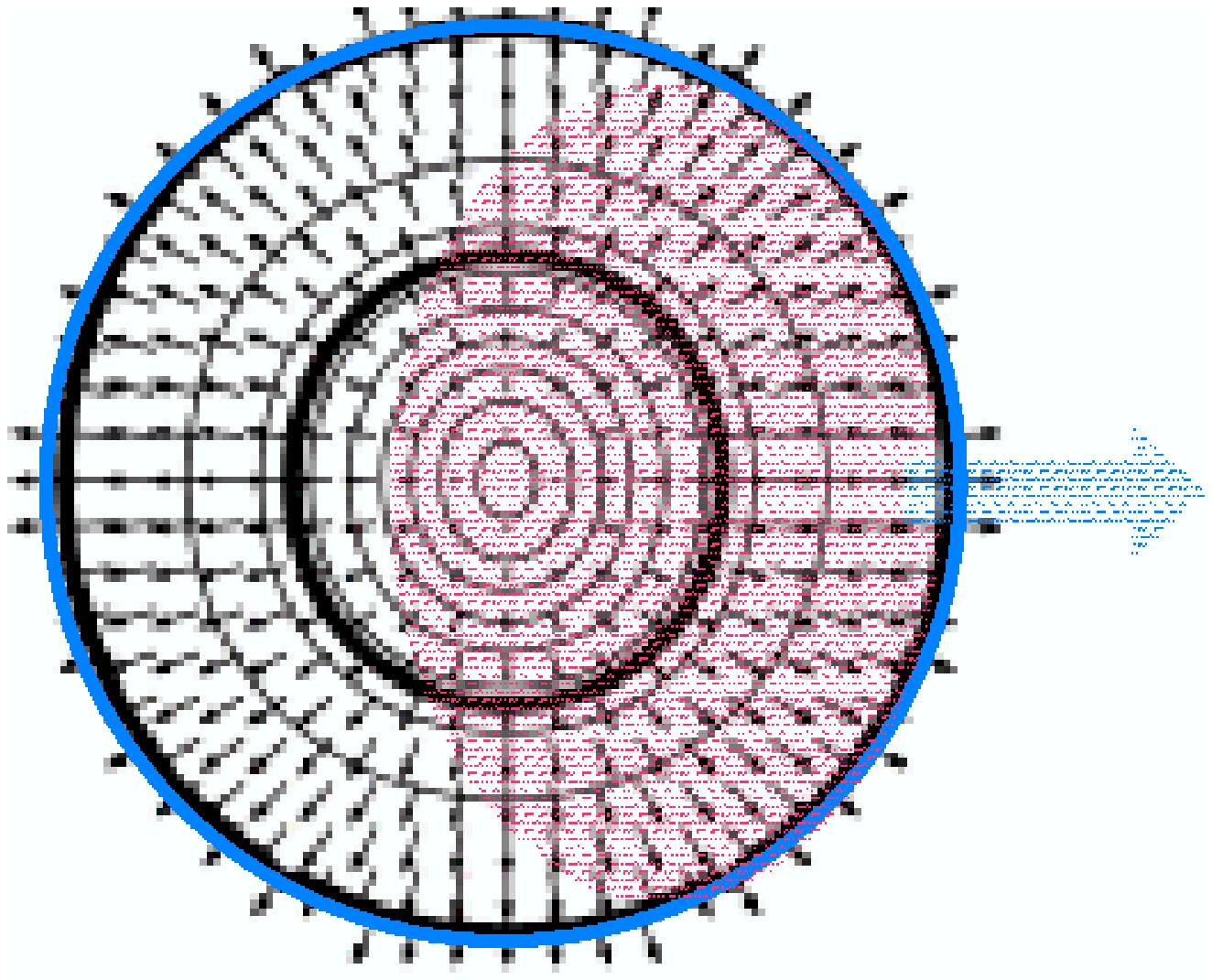}}}
 \end{minipage}
 \hspace{\fill}
 \begin{minipage}[t]{0.5\textwidth}
 {\centerline{\includegraphics[width=0.7\textwidth]{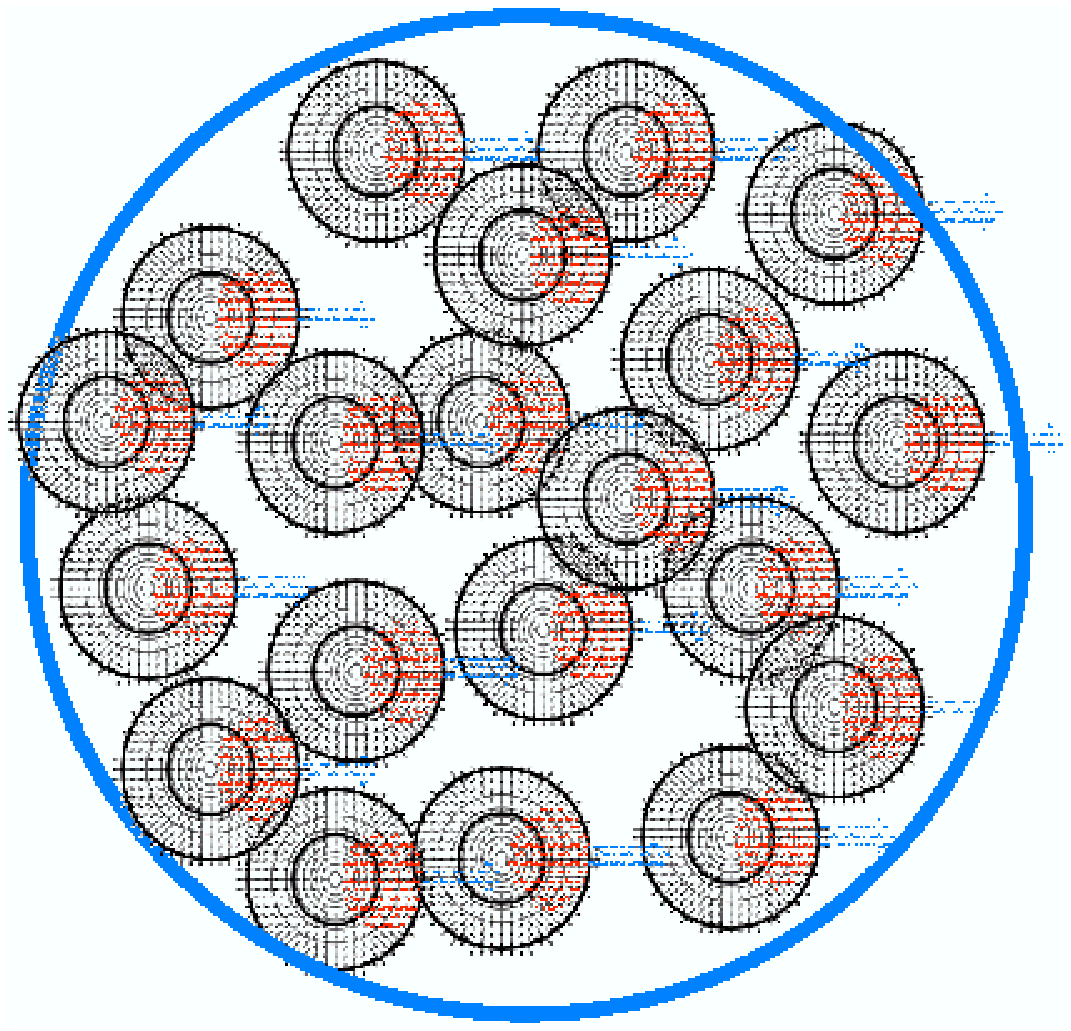}}}
 \end{minipage}
     \caption{Left: Global space-momentum correlations in a collectively-expanding system.  
 The homogeneity region for particles with a given $p_T$ is shaded.  Right: Local space-momentum correlations
 correspond to a ``super-''system composed of several {\it sub}-systems, each identical to that in the left panel.  The
 homogeneity region in this case is the sum of all the shaded areas-- that is, essentially the entire ``super-''system
 size.  While the geometric configurations of the two scenarios are quite different, they would appear identical
 in purely momentum space.}
     \label{fig:LocalAndGlobalXP}
 \end{figure}

HBT radius dependences on $p_T$ (c.f. Figure~\ref{fig:RLBWHBT}) are described by collective and thermal
dynamics as are momentum-space observables in flow-dominated ``Blast-wave'' models~\cite{RLBW}.  The
picture behind such models (which approximate hydrodynamical freeze-out scenarios) is illustrated in the
left panel of Figure~\ref{fig:LocalAndGlobalXP}.  There, arrows represent average boost velocities of local
elements of the emitting system.  Particles emitted with some $\vec{p}$ are preferentially emitted from one
region (called the ``homogeneity region''~\cite{Sinyukov}) of the source; higher-$p_T$ particles are emitted
from ever-smaller homogeneity regions.  
HBT radii reflect the size of this region.

In principle, temperature gradients~\cite{Csorgo} or a non-flowing but cooling source would also produce
{\it size}-momentum correlations and HBT radii which drop with $p_T$.  However, non-identical particle
correlations~\cite{STARkPi} show clear evidence for {\it position}-momentum correlations (note that the
average emission point depends on $\vec{p}$ in the left panel of Figure~\ref{fig:LocalAndGlobalXP}),
which arise only in the flow-dominated scenario.

\section{Puzzles, puzzles}
\vspace*{-2.0mm}

Into the comforting story described above, we inject some unpleasant facts.  Hydrodynamics--
the simple theory assuming only thermalization and using an equation of state--
does strikingly well in the description of the bulk momentum distribution of emitted particles.
However, hydrodynamical calculations fail to reproduce HBT radii.  Doubting the validity
of hydrodynamics or taking seriously purely hadronic dynamical models at RHIC~\cite{Tom} is
almost anathema.  Thus, the {\it first} ``puzzle'' is the failure of ``reasonable'' models
to reproduce the HBT radii.
It has been suggested that this puzzle is associated with timescales; the source evolution
and decoupling may occur more rapidly in Nature than in models.

While hydrodynamics-- even in the
momentum sector-- only works at RHIC (and, it is supposed, higher) collison energies, the
{\it second} femtoscopic ``puzzle'' arises from the violation of expectations based on very
general grounds.  As the collision energy is raised through the point at which QGP is barely
created, additional energy should go into the liberation of new degrees of freedom.  At this
point, the pressure would drop and expansion would slow due to the the ``softest point'' of
the Equation of State.
 This was expected to lead to a rapid rise in freezeout length- and time-scales,
measurable via femtoscopy.  Disapponting and puzzling indeed is the observation
of almost constant HBT radii as the energy scale is varied over two orders of magnitude~\cite{STARHBTPRC}.

When ``puzzles'' accumulate, it is well to revisit those aspects of the issue which
are assumed to be well-understood.  As mentioned above, the $p_T$-dependence of HBT radii
is thought to encode important dynamical information; the homogeneity lengths decrease with
increasing $p_T$ due to collective transverse flow.
Bulk flow requires substantial reinteraction of system components-- specifically,
a mean free path much smaller than the system size-- in order to develop.
Indeed, the failure of hydrodynamics to successfully describe data from lower energies
or peripheral collisions is attributed to insufficiently large matter density and/or
system size.
It might seem, then, that the collective flow signal in HBT radii from central Au+Au
collisions, would be smaller in peripheral collisions and smaller still in p+p collisions.

\begin{figure}[t!]
\begin{minipage}[t]{0.6\textwidth}
{\centerline{\includegraphics[width=0.95\textwidth]{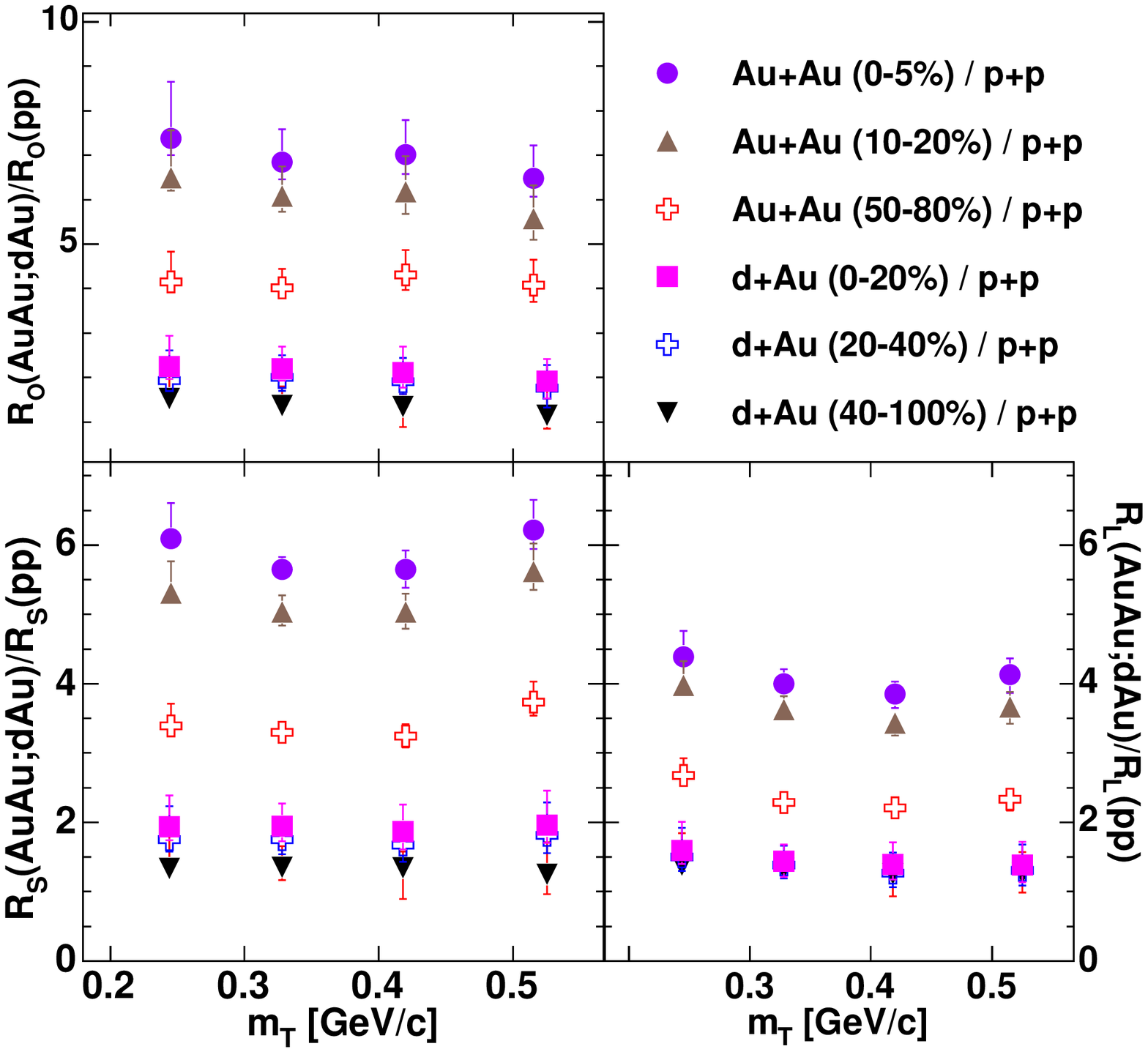}}}
    \caption{{\bf Preliminary} HBT radii from Au+Au and d+Au collisions with $\sqrt{s_{NN}}$ of various centralities have
been divided by those from p+p collisions at the same energy.  The ratio is plotted as a function of $p_T$.
}
    \label{fig:Division}
\end{minipage}
\hspace{\fill}
\begin{minipage}[t]{0.35\textwidth}
{\centerline{\includegraphics[width=0.95\textwidth]{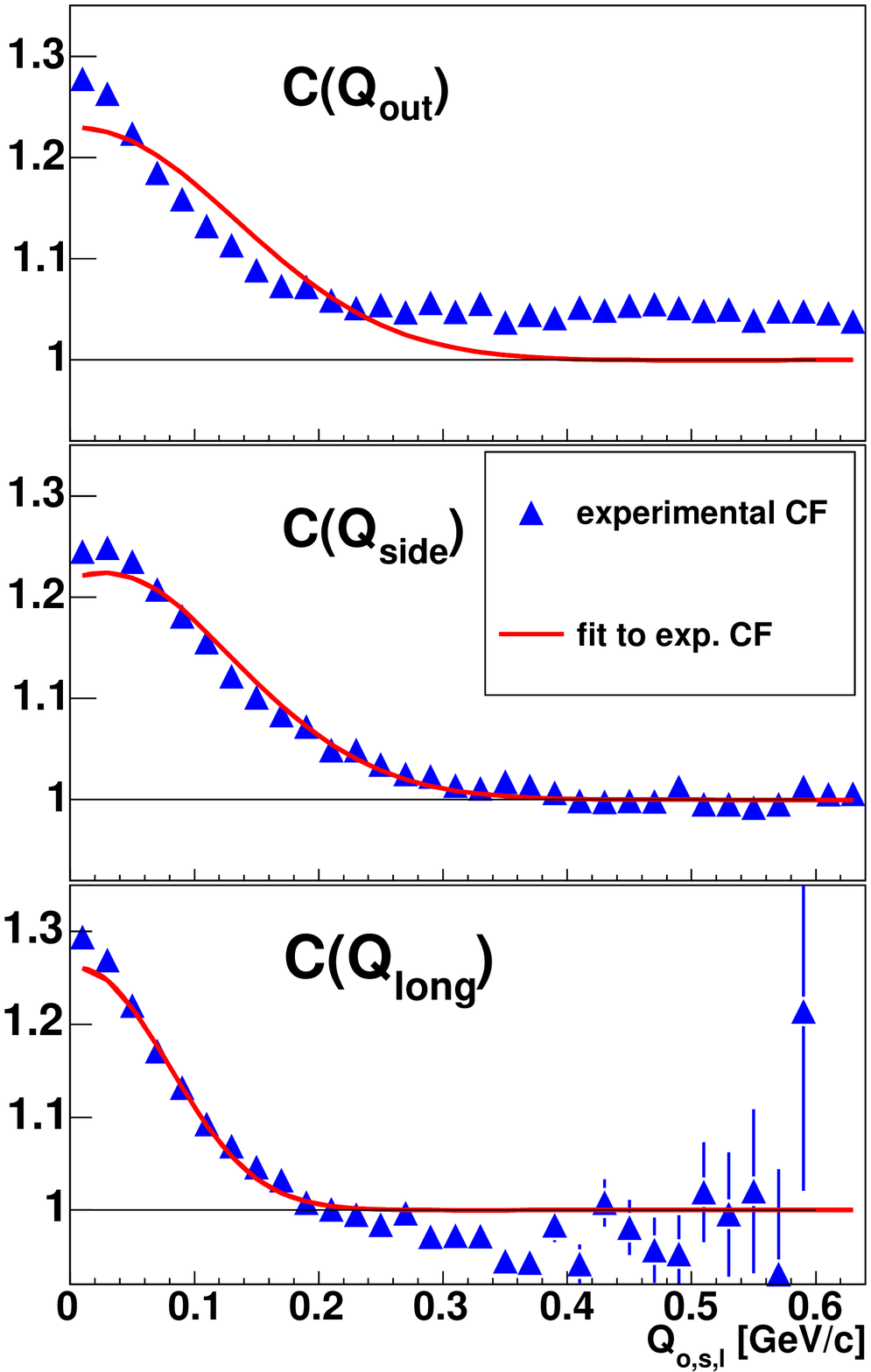}}}
    \caption{Cartesian (``out-side-long'') projections of {\bf preliminary} correlation functions
from peripheral d+Au collisions measured by STAR.
}
    \label{fig:dAuCartesian}
\end{minipage}
\end{figure}

In prior measurements~\cite{Alexander}, HBT radii from elementary collisions have been observed to decrease with $p_T$,
but this behaviour is generally
attributed to resonances or string dynamics.  STAR has also reported the
decrease with $p_T$ in preliminary HBT results in p+p collisions~\cite{TomsPoster}.  Given the rather
different underlying physics (bulk flow versus resonances or strings) behind the space-momentum correlations,
the observation~\cite{TomsPoster} of a trivial scaling $R^{p+p}(p_T) \sim R^{Au+Au}(p_T)$ is striking.
Figure~\ref{fig:Division} shows the ratio of HBT radii from Au+Au and d+Au collisions of various centrality,
divided by the HBT radii extracted from p+p collisions.  Although this scaling may be simply a benign coincidence,
for now we call this the {\it third} femtoscopic ``puzzle.''

\section{Viewing and diagnosing fine structure in 3D correlation functions}
\vspace*{-2.0mm}

As is commonly done, our HBT radii are extracted via fits to a functional form which assumes a Gaussian distribution in space
for the homogeneity region.  Not surprisingly, the distribution is never found to be {\it purely} Gaussian in practice,
and the consequences of non-Gaussianness (usually within a model-dependent context) have been discussed
extensively in the literature.  One of the few quantitative experimental evaluations of systematic
errors associated with non-Gaussian components has recently been published~\cite{STARHBTPRC}.  There, we
conclude that while such components are significant, the Gaussian HBT radii do characterize the correlations,
even when a more general functional form is used.  Here, we discuss a different aspect of the measured correlations
for the lightest systems.

\subsection{Problems in (and with) Cartesian projections}

Correlation functions are measured and fit in three-dimensional relative momentum components
$q_{o}-q_{s}-q_{l}$ (``out-side-long'') space~\cite{BP} using
a functional form~\cite{STARHBTPRC}
\begin{equation}
\label{eq:GaussForm}
C(q_{o},q_{s},q_{l}) = \frac{A(q_{o},q_{s},q_{l})}{B(q_{o},q_{s},q_{l})} =
N\cdot\left[(1-\lambda) + \lambda K_{coul}(q_{inv})\left( 1+e^{-q_{o}^2R_{o}^2-q_{s}^2R_{s}^2-q_{l}^2R_{l}^2} \right) \right]
\end{equation}
where $K_{coul}(q_{inv})$ is the squared relative wavefunction integrated over a Gaussian source of finite size, and $N$
is a normalization factor independent of $\vec{q}$.

Visual evaluations of the quality of the data and fit are typically done via 1-dimensional projections 
 of the data and the fit onto $q_i$ ($i=o,s,l$):
\begin{equation}
\label{eq:CartesianProjections}
C^i_{\rm exp}(q_i) = \frac{\sum_{j,k} A(q_{o},q_{s},q_{l})}{\sum_{j,k} B(q_{o},q_{s},q_{l})} \quad \quad
C^i_{\rm fit}(q_i) = \frac{\sum_{j,k} T(q_{o},q_{s},q_{l}) \cdot B(q_{o},q_{s},q_{l})}{\sum_{j,k} B(q_{o},q_{s},q_{l})} .
\end{equation}
Here, $T(q_{o},q_{s},q_{l})$ represents the right-hand side of Equation~\ref{eq:GaussForm}.  Sums in the projection for each
component $i$ are done over bins with $q_{j}$ and $q_{k}$ ($k,j \neq i$) in some small range-- typically on the scale of
the femtoscopic correlation enhancement.  Equation~\ref{eq:CartesianProjections} defines what we call ``Cartesian projections''
in the out-side-long space.

Figure~\ref{fig:dAuCartesian} shows such projections for preliminary correlation functions for peripheral $d+Au$ collisions.
Clearly, the fit and data projections in $q_{o}$ are significantly different in the femtoscopic (low-$q$) region.  To first
order, this suggests that the extracted $R_{o}$ is too low by $\sim 30 \%$.  However, this is not the full story, as we do
not see the full 3D structure of the correlation functions in projections from Equation~\ref{eq:CartesianProjections}; it may
be that nontrivial structure in {\it un}plotted $q$-bins cause the fit (which {\it globally} minimizes some figure of merit
such as negative-log-likelihood) to make compromises which show up in this particular projection.  Similarly,
non-experts sometimes scratch their heads when viewing 1-dimensional projections: there is often no {\it apparent} reason
why the fit should not do much beter than it does.

In Figure~\ref{fig:dAuCartesian}, however, we see clear problematic structure even in the Cartesian projections for d+Au data.
In particular, the correlation function in the large-$q$ (``normalization'') region has different values in the different 
projections.  
Pions emitted from p+p collisions have a very large Bohr radius compared to the source size, so
structure at very large $q$ cannot be of femtoscopic origin as quantum statistics-induced correlations
die out, and Coulomb final-state interactions will be overwhelmingly s-wave.
It could in principle arise due to
global or other correlative effects such as momentum conservation or jets, though our preliminary studies suggest that
this is not the case here.  We do not present these studies or further discuss possible origins of the structure, but
only focus on its characterization here.  This preliminary result in Figure~\ref{fig:dAuCartesian} using low-multiplicity
data from STAR represents the worst-case dataset, and so is a good dataset on which to focus efforts to
study nontrivial structure in the correlation functions and the extent to which they influence the extracted femtoscopic scales.

\subsection{Representation via spherical decomposition}

In principle, one could visualize the full structure of the 3-dimensional correlation function via a {\it series} of Cartesian
projections in $q_i$ over different ranges in $q_{j,k}$.  This would, however, constitute a large number of figures,
and relevant patterns which cut across projections might not catch the eye.  Given the symmetries of the correlation in
$\vec{q}$-space, a more natural representation is through spherical decomposition.

Here, the spherical coordinates $\theta$, $\phi$, and $Q$ relate to the Cartesian ones as
\begin{equation}
q_{o} = Q\sin{\theta}\cos{\phi} , \qquad 
q_{s} = Q\sin{\theta}\sin{\phi} , \qquad 
q_{l} = Q\cos{\theta} .
\end{equation}

The correlation function may be binned\footnote{To avoid artifacts,  bin widths should be $\Delta_\phi = \frac{2\pi}{N_\phi}$
and $\Delta_{\cos\theta} = \frac{2}{N_{\cos{\theta}}}$.}
in $Q$, $\cos{\theta}$, and $\phi$ and represented via its spherical harmonic coefficients, which will
depend on Q:
\begin{equation}
\AlmQ = \sum_{{\rm bins} \thinspace i} C\left(Q,\theta_i,\phi_i \right) Y_{\lambda,\mu}\left(\theta_i,\phi_i\right) 
                                       F_{\lambda,\mu}\left(\theta_i,\Delta_{\cos\theta},\Delta_\phi\right) ,
\end{equation}
where $F_{\lambda,\mu}$ represents a numerical factor correcting for finite bin sizes $\Delta_{\cos\theta}$ and $\Delta_\phi$; 
it turns out not to depend on $\phi_i$.

Symmetry constrains the number of relevant components.
For femtoscopic analyses of identical particles at midrapidity which integrate over reaction-plane orientation (i.e. almost all analyses to date), only 
\Alm's with even values of $\lambda$ and $\mu$ do not vanish.  Further, it is natural to expect that the statistical relevance of 
high-$\lambda$ components is diminished.  Thus, by glancing at only a {\it few one-dimensional plots}, one views the {\it entire} correlation
structure in orthogonal components.  Once the practitioner is familiar with this representation, it should have diagnostic value.

Another advantage of the spherical representation is that they are less sensitive to detector acceptance effects than are the
Cartesian projections.  While the {\it three}-dimensional correlation function is ideally independent of acceptance, the summing
and weighting with bins in Equation~\ref{eq:CartesianProjections} will affect $C^i(q_i)$.  Thus, while NA44 and NA49 should measure
the same correlation function in $\vec{q}$-space, their 1-dimensional projections will differ.
Their \Alm's, however, will be identical, assuming that both experiments have {\it some} nonvanishing acceptance in all $\vec{q}$ bins.
Related to this, the one-dimensional correlation function $C(Q_{inv})$ depends heavily on acceptance, while $A_{0,0}(Q)$ does not.
Finally, from a physics standpoint, it has recently been pointed out~\cite{DaniPratt} that \Alm ~components
correspond one-to-one with anisotropies of the same order $\lambda, \mu$ of the spatial homogeneity region.

\begin{figure}[h!]
{\centerline{\includegraphics[width=0.9\textwidth]{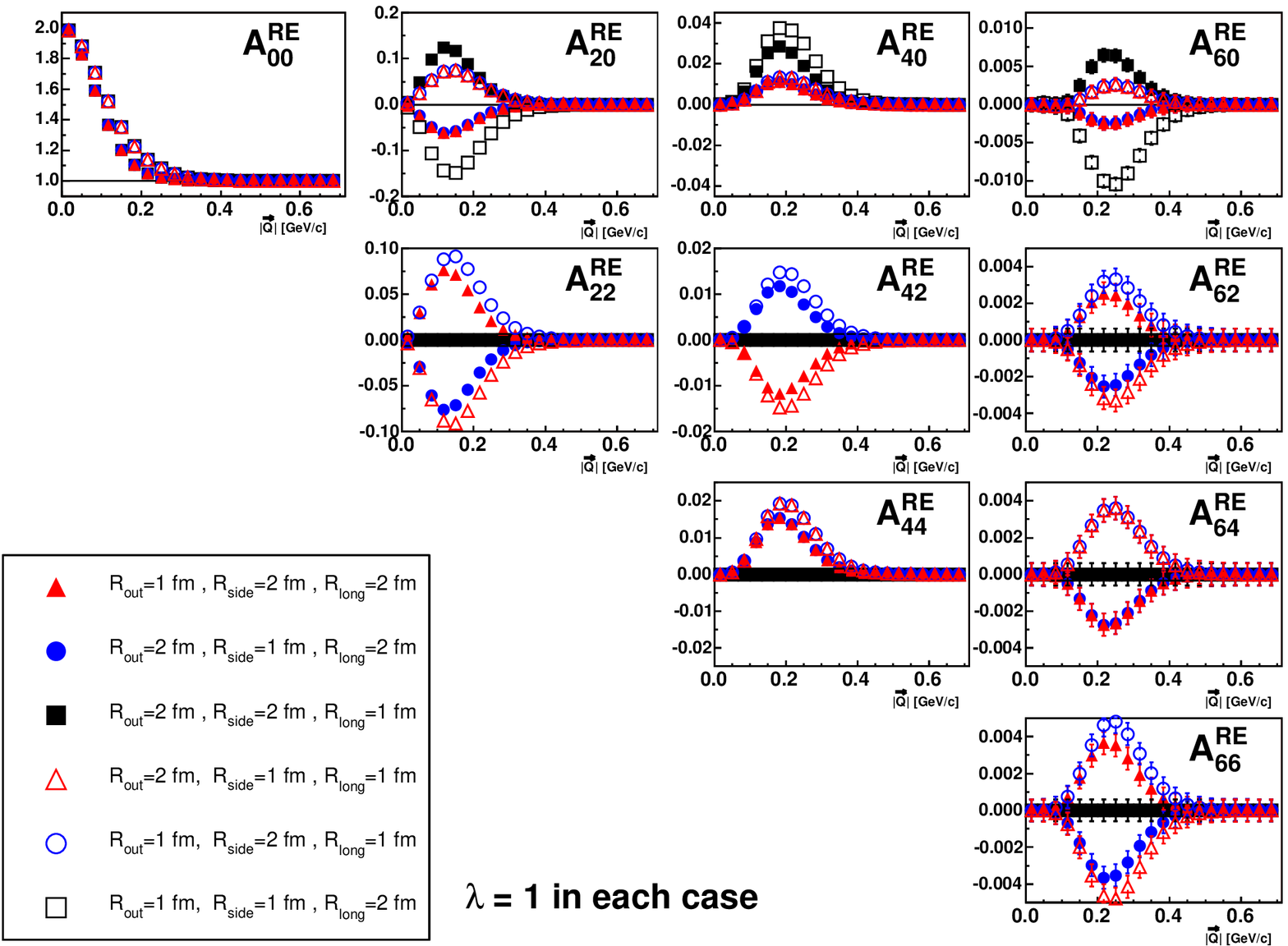}}}
    \caption{Spherical harmonic coefficients \AlmQ~ corresponding to a purely Gaussian correlation
function with various HBT radii.
}
    \label{fig:AlmModel}
{\centerline{\includegraphics[width=0.9\textwidth]{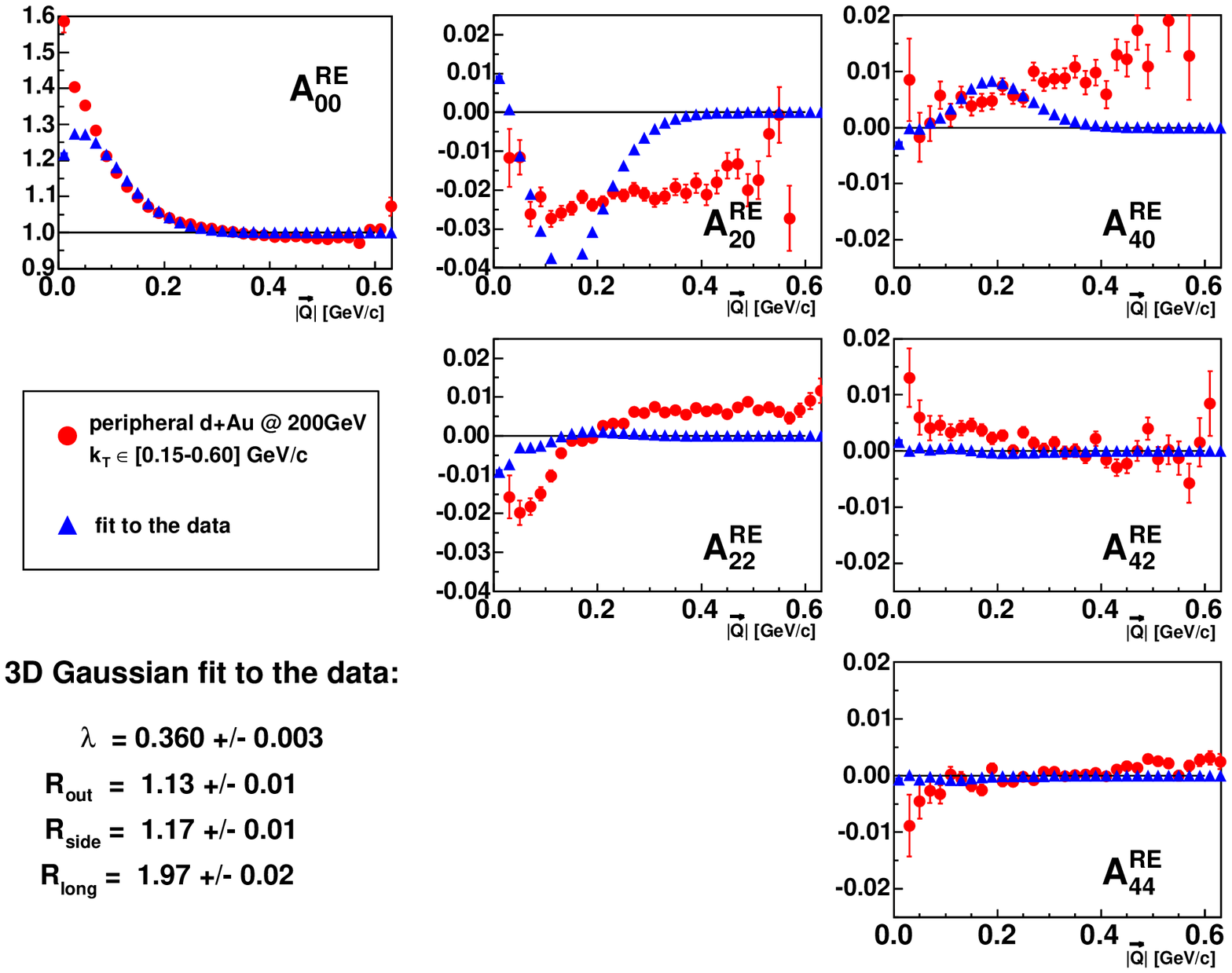}}}
    \caption{{\bf Preliminary} spherical harmonic coefficients \AlmQ~ from peripheral d+Au collisions,
together with the harmonics from a Gaussian fit.
}
    \label{fig:AlmData}
\end{figure}

To begin calibrating ourselves in this new representation, we show in Figure~\ref{fig:AlmModel} harmonic coefficient functions corresponding to a purely
Gaussian homogeity region.  (In this simple example, $K_{coul}$ in Equation~\ref{eq:GaussForm} has been set to unity.)
Some simple intuitive trends are clear.  For this case, $A_{0,0}(Q)$ depends only on $R^2_o+R^2_s+R^2_l$ and not on shape information.
The sign and magnitude of $A_{2,0}$ reflects the relative magnitude of the longitudinal to the transverse radii, independent of $R_o/R_s$.
In contrast, the important comparison of $R_o$ and $R_s$ is reflected in the sign of $A_{2,2}$.  Information encoded in higher-order components 
is more complicated, but clearly simple degeneracies and sign-changes appear there, as well.

Generic features are also clear.  We notice further that, with increasing $\lambda$, the ``bump'' in \AlmQ becomes increasingly smaller in
magnitude, as hoped (so that only a few $\lambda$ terms are relevant).
Also, $\AlmQ \stackrel{Q \to 0}{\rightarrow} 0$ if $\lambda \neq 0$, and the ``bump'' moves to higher $Q$ with increasing $\lambda$, reminiscent
of wavefunctions for central potentials.
Finally, the calculated correlation function approaches a constant value (here, unity) for large $Q$, independent
of direction ($\theta,\phi$); thus, $\Alm\stackrel{Q\to\infty}{\rightarrow}\delta_{0,\lambda}$.

The spherical decomposition for the worst-case (low multiplicity) d+Au data is shown in Figure~\ref{fig:AlmData},
together with the decomposition of the fit.  \Alm ~components for $\lambda > 4$ were statistically insignificant.
Clearly, the functional form in Equation~\ref{eq:GaussForm} (now with the Coulomb term) does not describe the
correlation anisotropy or strength.
The overall size probed by $A_{0,0}$ is probably accurately extracted, but the anisotropic
shape (e.g. $R_o/R_s$) may be underestimated in the fit because of the influence of large-$Q$ correlations.
The nonfemtoscopic $\lambda \neq 0$ components appear roughly independent of $Q$, and might contribute as well
into the femtoscopic (low-$Q$) region.
  At present, we are actively investigating the source of the large-$q$ anisotropy in the data
and how to account for it in the fit.

\section{Summary}
\vspace*{-2.0mm}

We have presented preliminary HBT radii for d+Au and p+p collisions and pointed out a surprising scaling effect when they
are compared with those published for Au+Au collisions.
A naif interpreting Figure~\ref{fig:Division} would conclude that the linear scale of a gold nucleus is about six times
that of a proton; despite the tremendous detail accummulated since then, this threatens to return femtoscopy to its
status of 15 years ago (c.f. Figure~\ref{fig:Status15YearsAgo}).
With arguments of flow and space-momentum correlations and 
expansion\footnote{The ratio 6:1 between the HBT radii (R.M.S. values at freezeout) for Au+Au versus p+p collisions should probably be
compared to a 3:1 ratio between the R.M.S. of Au and proton.
Thus, the constant ratio $\sim 197^{1/3}$ might be coincidental, masking greater  expansion in the Au+Au case.},
together with a dose of coincidence,
a more sophisticated expert would claim that much more information is encoded in the HBT systematics.  However, 
Occam's razor is a formidable adversary.  It is urgent to study the ``third femtoscopic puzzle'' in detail.

We are exploring a new representation of the multidimensional correlation function which has several technical advantages
and a simple connection to physics.  This
representation makes visibly manifest the full structure of the correlations, and has revealed, for the small
collision systems, long-range nonfemtoscopic correlations of unknown origin.  At present, it is unclear the extent
to which these not-understood features are related to the ``third puzzle.''  Model studies into their origin, and
attempts at more general fits to the data, are underway.


\vfill\eject
\end{document}